\documentclass[11pt, a4paper]{article}

\usepackage{amsmath}
\usepackage{amsfonts}
\usepackage{amssymb}

\date{}
\begin{document}

\title{Gravity as a gauge theory of translations}
\author{J.  Mart\'{\i}n--Mart\'{\i}n \thanks{Departamento de F\'{\i}sica
Fundamental,  Universidad de Salamanca,  37008 Salamanca,  Spain, 
e-mail: chmm@usal. es} \and A.  Tiemblo\thanks{Instituto de F\'{\i}sica Fundamental,  C/ Serrano 113 bis, CSIC,  28006, Madrid,  Spain, 
e-mail: Tiemblo@imaff.cfmac.csic.es}}

\maketitle

\begin{abstract}
The Poincar\'e group can be interpreted as the group of isometries
of a minkowskian space.  This point of view suggests to consider the
group of isometries of a given space as the suitable group
to construct a gauge theory of gravity.  We extend these ideas to
the case of maximally symmetric spaces to reach a realistic theory
including the presence of a cosmological constant.  Introducing the
concept of ``minimal tetrads" we deduce Einstein gravity in the
vacuum as a gauge theory of translations. 
\end{abstract}

\section{Introduction}
Universality is,  without any doubt,  the characteristic property of
gravity,  a force affecting in the same manner to all kinds of
matter and energy.  It was probably this universality the key idea
leading Einstein to identify gravity with a very fundamental
ingredient of reality.  According to the contemporary mathematical
achievements,  he found in Geometry the final answer.  Coherently
with this basic assumption,  our description of the gravitational
field is given by some geometrical background perturbed by the
presence of matter or energy. 

As a matter of fact our phenomenological experience of the
space--time nature is,  roughly speaking,  very close to an almost
pseudoeuclidean geometry slightly modified by the presence of
masses or energy.  Here the term slightly alludes to the
experimental evidence of the extreme weakness of gravity when
compared to the other natural forces.  These properties are on the
basis of the so-called ``weak field approximation" which
constitutes an effective approach well adapted to many different
problems.  In any case the idea that gravity could be described by
some expansion involving a characteristic parameter must be,  in our
opinion taken seriously into account.  A part of this paper is
precisely devoted to investigate the possible theoretical grounds
of such an expansion. 

In the absence of fermions the theory can be established in a
purely geometrical framework.  Nevertheless,  a relevant question
arises when we attempt to couple fermions with gravity,  a program
which implies the necessity to enlarge the framework with the
introduction of an internal symmetry group suitable to include
spinorial fields in the theory. 

There is, in fact, a very rich
literature \cite{Utiyama} \cite{Sciama} \cite{Kibble}
 \cite{Hayashi} \cite{Ivanenko} \cite{Lord1} \cite{Lord2} \cite{Sardanashvily} \cite{Ashtekar} suggesting different extensions of the gauge principle to space--time
groups,  an attempt that exploits the properties of the Yang--Mills
Theories in the hope to express gravity as mediated by gauge
potentials only,  as it happens with the other interactions. 

Special mention is deserved to the occurrence in these kind of
theories of the tetrad a typical structure which is neither a
purely geometrical object nor an internal tensor,  but both things
at the same time,  as deduced from the double behavior with respect
to space-time coordinates and internal degrees of freedom.  To
explain tetrad's properties several proposals have been considered
and to this purpose Hehl's Poincar\'e gauge theory as well as
metric affine gravity \cite{Hehl1} \cite{Hehl2} \cite{Hehl3} must be
mentioned. 

It must be stressed that tetrads can be considered as the
footprint of the unavoidable presence of translations in a true
gauge theory of gravity.  Using Feynman's words ``gravity is that
field which corresponds to a gauge invariance with respect to
displacement transformations".  We claim that the cornerstone to
include translations in such a gauge theory is given by non linear
realizations of space-time symmetry groups containing
translations \cite{Coleman} \cite{Callan} \cite{Salam} \cite{Isham} \cite{Borisov} \cite{Cho} \cite{Stelle}.  
On these grounds we are going to show that the formal structure of
tetrads can be expressed in terms of well known gauge objects, 
 connections and Goldstone--like bosons associated to the
displacement transformations that,  as we will see,  can be locally
chosen as coordinates that,  in this way,  can be interpreted as well
as dynamical objects. 

We devote the first section to a brief review of the local non
linear realizations.  The application to the Poincar\'e group is
also included,  obtaining the formal structure of the tetrads and
discussing the possible dynamical interpretation of its different
elements. 

Being the quantum vacuum the basic ingredient of the physical
reality,  any realistic attempt to construct a dinamical
description of the space--time nature,  must include the role of the
cosmological constant as an initial requirement.  In this aim the
extension of the gauge principle to the group of isometries of a
space of constant curvature,  appears as the natural generalization
of the Poincar\'e group that can be considered as a limiting case
when the curvature tends to zero. 

The procedure provides a structure for the tetrads that allows us
to identify the minimal structure satisfying all the requirements
demanded for a tetrad.  Curiously the remaining part is associated
only to the translational connections that,  in this manner,  can be
interpreted,  following Feynman ideas,  as the true gravitational
contributions. 

On these assumptions a decomposition,  analogous to the ``weak field
approach" is naturally obtained,  it nevertheless appears not as a
reasonable approximation but as an exact structure predicted by
the non linear realization of the symmetry group. 

\section{Non linear tetrads}
To construct a gauge theory of gravity,  one is constrained to start
from our phenomenological evidence of the space--time observable
properties.  Therefore,  being Poincar\'e the group of
isometries of the Minkowskian space,  the presence
of translations in the gauge description is a necessary
consequence. 

The non linear gauge realizations of space--time symmetry groups
containing translations have been the object,  in the past,  of
several papers,  \cite{Julve1} \cite{Julve2} \cite{Lopezpinto} \cite{Lopezpinto2}
 \cite{Tresguerres}.  Nevertheless to facilitate the reading of this work
we include here a brief review of the methods and main results. 

Let $G$ be a Lie group having a subgroup $H$,  we assume that the
elements $C(\varphi)$ (cosets) of the quotient space $G/H$ can be
characterized by a set of parameters say $\varphi$.  Let us denote
by $\psi$ an arbitrary linear representation of the subgroup $H$. 

The non linear realization can be deduced from the action of a
general element  $``g"$ of the whole group on the cosets
representatives defined in the form:

\begin{equation}
\label{1}
g C(\varphi)=C(\varphi')h(\varphi, g)
\end{equation}
where $h(\varphi, g)\in H$. It acts linearly on the representation
space $\psi$ according to:

\begin{equation}
\label{2}
\Psi'=\varrho[h(\varphi, g)]\Psi, 
\end{equation}
being $\rho$ a representation of the subgroup $H$.  It can also be
seen that the coset realization is the most general one which
preserves linear the action of the subgroup $H$. 

To construct a non linear local theory,  the next step is to define
suitable gauge connections,  that can be obtained by substituting
the ordinary Cartan 1--form $\omega=C^{-1}dC$ by a generalized
expression of the form:

\begin{equation}
\label{3}
\Gamma=C^{-1}DC
\end{equation}
where $D=d+\Omega$ is the covariant differential constructed with
the $\Omega$ 1-form connection defined on the algebra of the whole
group and having the canonical transformation law:

\begin{equation}\label{4}
\Omega'=g\Omega g^{-1}+  gdg^{-1}
\end{equation}

The main point to be emphasized concerns the components of the non
linear connection $\Gamma$ that can be classified into two
different categories
\begin{itemize}
     \item Those associated to the subgroup $H$,  having the character
     and transformation properties of an ordinary gauge connection
     with respect to the action of $H$. 

     \item Those associated to the coset sector,  having the meaning of
     covariant differentials of the coset fields $\varphi$ and
     transforming as a true tensor with respect to the subgroup
     $H$. 
\end{itemize}

The method can be applied to the Poincar\'e group \cite{Martin} by
choosing $H= Lorentz$ and parametrizing the cosets in the form
$C(\varphi)=e^{i\varphi^iP_i}$,  where $P_i$ are the translational
operators and $\varphi_i$ a set of continuous parameters. 

Using \eqref{1} a simple calculation yields the variations of the cosets
parameters $\varphi$ which reads:
\begin{equation}\label{5}
\delta\varphi^i=\epsilon^i+\beta^i_j\varphi^j
\end{equation}
 where $\beta^i_j$ are the infinitesimal parameters of the
 Lorentz transformations and $\epsilon^i$ the corresponding ones
 for the translations. 

 The general non linear connection \eqref{3} contains the ordinary
 Lorentz linear gauge connection $A_{ij}$ and a Lorentz tensorial
 object $e_i$ linked to the coset sector that can be identified with the
 tetrad,  namely:
\begin{equation}\label{6}
e_i=D\varphi_i+\Gamma(T)_i
\end{equation}
where D is the Lorentz covariant differential and ${\Gamma(T)_i}$
is a translational gauge connection transforming as
\begin{equation}\label{7}
\delta\Gamma(T)_i=\beta_i^j\Gamma(T)_j-D\varepsilon_i
\end{equation}. 

According to the transformation properties \eqref{5} the fields
$\varphi^i$ can be identified with the cartesian coordinates,  a
nice result as long as in this way they can be dynamically
interpreted as Goldstone bosons with respect to the gauged
translations.  A second question arises from the anomalous
dimensionality of the translational connection $\Gamma(T)$ in
equation \eqref{6}. Being a connection  an object with the same formal dimensionality of a derivative,  we introduce a constant
characteristic length, say $\lambda$,  to render it homogeneous with
the ordinary Lorentz connection $A_{ij}$.  Thus redefining
$\Gamma(T)_i=\lambda\gamma_i$ we rewrite explicitly:
\begin{equation}\label{8}
e(\lambda)_{\mu i}=\partial_\mu\varphi_i+A_{\mu i
j}\varphi^j+\lambda\gamma_{\mu i}
\end{equation}

The occurrence of a fundamental length in gravitational physics is
an almost commonly accepted idea,  it appears for instance at the
Planck scale,  lattices or string
theories,  \cite{Borzeskowski} \cite{Garay} \cite{Berg} \cite{Feinberg}
 \cite{Kato} \cite{Konishi}.  We claim that a natural place to include
a fundamental length is precisely the translational connection in
equation \eqref{6},  as we will see the smallness of $\lambda$ gives a
coherent support to the treatment of gravity as a perturbation of
a background metric. 

As it has been pointed out,  the essential property of a tetrad is
given by its double character,  transforming as an ordinary general
vector with respect to coordinates transformations in the Greek
indices and as a Lorentz vector in the Latin ones.  It acts,  in this
way,  as a link between both spaces.  Nevertheless from equation 
\eqref{8} we realize that,  being a covariant derivative of a Lorentz
vector,  $e(0)_{\mu i}=\partial_\mu\varphi_i+A_{\mu i j}\varphi^j$
is the minimal structure suitable by itself to accomplish this
function,  we outline that the distinction between $e(0)_{\mu i}$
and $e(\lambda)_{\mu i}$ depends only on the behavior with respect
to translations. 

Now with the help of the general tetrad \eqref{8} we define,  in the usual way,  the metric tensor
\begin{equation}\label{9}
g(\lambda)_{\mu \nu}=e(\lambda)_\mu^i e(\lambda)_{\nu i}=g(o)_{\mu
\nu}+\lambda\gamma_{\left\{\mu \nu\right\}}+\lambda^2\gamma_{\mu
\rho}\gamma_{\nu \sigma}g(0)^{\varrho \sigma}
\end{equation}
where $g(0)_{\mu \nu}=e(0)_{\mu i}e(0)_\nu^i$ and we have used
$e(0)_{\mu i}$ to transform indices. 

Equation \eqref{9} strongly resembles a weak field expansion,  playing
$g(0)_{\mu \nu}$ the role of a background metric,  nevertheless it
must be emphasized that equation \eqref{9} is an exact result deduced
from the structure of the tetrads.  Therefore to elucidate the properties,
 structure and dynamical nature of $g(0)_{\mu \nu}$ becomes a
relevant and interesting question which is going to be
investigated in what follows. 

As we have pointed out the choice Poincar\'e as the internal group
is based on our phenomenological knowledge of space--time which
suggests us a configuration very close to a minkowskian
space. Notwithstanding a pure psudoeuclidean model excludes the
presence of a cosmological constant, a contribution that describes
essential properties of the quantum vacuum or, in other words, of
the real physical space-time. Therefore being Minkowsky the
simplest case of the spaces of constant curvature it seems natural
to extend the study to the general case to include the possible
occurrence of a cosmological constant. The relevance of this point
will be evident in the next section. 

A maximally four dimensinoal symmetric space, in fact, admits a maximal number of
Killing vectors defined as:
\begin{equation}
\hat{p_i}=i \left\{\partial_i+\frac{k}{4}(2x_ix^j-\delta_i^j
{r}^2)\partial_j\right\}
\end{equation}
\begin{equation}
L_{ij} =i(\delta_i^kx_j-\delta_j^kx_i)\partial_k
\end{equation}
where ${r}^2=\eta_{ij}x^ix^j$ and being k the curvature.  They
define a semisimple Lie algebra of the form:
\begin{equation}
[\hat{p_i}, \hat{p_j}]=ikL_{ij}
\end{equation}
\begin{equation}
[L_{ij}, \hat{p_k}]=i\eta_{k[i}\hat{p}_{j]}
\end{equation}
\begin{equation}
[L_{ij}, L_{kl}]=-i\{\eta_{i[k}L_{l]j}-\eta_{j[k}L_{l]i}\}
\end{equation}
which reduces to Poincar\'e when $k\rightarrow 0$. 

To construct the local theory we follow the general scheme
outlined for the Poincar\'e case. Therefore we maintain for the
cosets the same definition taking
$C=e^{i\varphi^i\hat{p_i}}$, parametrizing the group element $g$
and the elements of $H=Lorentz$ respectively as:
\begin{equation}
g=e^{i\epsilon^i\hat{p_i}}e^{\frac{i}{2}\beta^{ij}L_{ij}}\thickapprox
I+i(\epsilon^i\hat{p_i}+\frac{1}{2}\beta^{ij}L_{ij})
\end{equation}
and
\begin{equation}
h=e^{\frac{i}{2}u^{ij}L_{ij}}
\end{equation}
where $\epsilon^i$ and $\beta^{ij}$ are the corresponding
infinitesimal parameters. 

The non linear expression of the connections can be deduced from
the general formula \eqref{3} which in this case reads out:
\begin{equation}
\Gamma=C^{-1}DC=e^{-i\varphi^i\hat{p_i}}[d+i\lambda\gamma^i\hat{p_i}+\frac{i}{2}A^{ij}L_{ij}]e^{i\varphi^i\hat{p_i}}
\end{equation}
With the help of the commutation relations (12), (13), (14) and
using Hausdorff--Campbell formulas to deal with exponentials we
obtain, after a little calculation,  the value of $\Gamma$
\begin{equation}
\Gamma=i\hat{e}_i\hat{p}^i+\frac{i}{2}\hat{A}_{ij}L^{ij}
\end{equation}
with
\begin{equation}
\hat{e}_i=ND\varphi_i+\frac{1-N}{\mu^2}(\varphi^jD\varphi_j)\varphi_i+\lambda
M\gamma_i+\lambda\frac{1-M}{\mu^2}(\gamma^j\varphi_j)\varphi_i
\end{equation}
where $D$ is the Lorentz covariant
differential, $\mu=\varphi_i\varphi^i$ being $M$ and
$N$, respectively:
\begin{equation}
M=1-\frac{k\mu^2}{2!}+\frac{(k\mu^2)^2}{4!}\cdots \thicksim
cos\sqrt{k\mu^2}
\end{equation}
\begin{equation}
N=1-\frac{k\mu^2}{3!}+\frac{(k\mu^2)^2}{5!}\cdots \thicksim
\frac{sen\sqrt{k^2}}{\sqrt{k\mu^2}}
\end{equation}
and
\begin{equation}
\hat{A}_{ij}=A_{ij}+\frac{1-M}{\mu^2}\varphi_{[i}D\varphi_{j]}+\lambda
N\varphi_{[i}\gamma_{j]}
\end{equation}

To use only non linear connections we introduce in (19) the value
of $A_{ij}$ as deduced from (22),  so that we rewrite (19) in the
form:
\begin{equation}
\hat{e}_i=\frac{N}{M}\hat{D}\varphi_i+\frac{1}{\mu^2}(1-\frac{N}{M})(\varphi^jd\varphi_j)\varphi_i
+\lambda\frac{1}{M}\gamma^*_i+\lambda\frac{\varphi_i\varphi^j}{\mu^2}\gamma_j
\end{equation}
where $\hat{D}$ means the Lorentz covariant differential in terms
of $\hat{A}_{ij}$, and $\gamma^*_i=(\delta_{ij}-\varphi_i\varphi_j/\mu^2)\gamma^j$. 

Using the same techniques the non linear transformations
properties of the fields $\varphi_i$ can be deduced from the non
linear action definition (1), namely:
\begin{equation}
\delta\varphi_i=\frac{1}{N}\left[\varepsilon_i+(N-1)\frac{(\varphi^j
\epsilon_j)\varphi_i}{\mu^2}\right]+u_{i j}\varphi^j
\end{equation}
with:
\begin{equation}
u_{i j}=\beta_{i j}+\frac{1-M}{N \mu^2}\varphi_{[ i}\varepsilon_{j
]}
\end{equation}

It can be easily seen that for a vanishing $k$ we recover the
Poincar\'e approach as given by (8). Of course the limit $
K\rightarrow0$ must be considered carefully, they are in fact
different group structures with specific properties. The Cassimir
operator, for instance have the expression
$C=\frac{1}{6K}\triangle$ which depends on the inverse of $K$.

 A redefinition of the form $\hat\varphi_i=\varphi_iN$ leads us
finally to an alternative more compact expression for $\hat{e_i}$, 
namely:
\begin{equation}
\hat{e}_i=\frac{1}{\hat{M}}(\hat{D}\hat\varphi_i+\lambda\gamma^*_i)+\lambda\frac{\hat\varphi_i\hat\varphi_j}{\hat\mu^2}\gamma^j
\end{equation}

To conclude this section we explicit write $\hat{e}_i$ in its
tensorial form as:
\begin{equation}
\hat{e(\lambda)}_{\mu i}=\hat{e(0)}_{\mu i}+\lambda
\hat{\gamma}_{\mu i}=\hat{e(0)}_{\nu i}(\delta^\nu_\mu+\lambda
\hat{\gamma}^\nu_\mu)
\end{equation}
where
\begin{equation}
\hat\gamma_{\mu i}=\frac{1}{\hat{M}} {\gamma}_{\mu
i}\ast+\frac{\varphi_i \varphi_j}{\hat\mu^2} \gamma^j_\mu
\end{equation}
and
\begin{equation}
\hat{e(0)}_{\mu i}=\frac{1}{\hat{M}} \hat{D}_\mu \hat{\varphi}_i
\end{equation}

In this manner the theory can be interpreted as depending on two
different dynamical variables, a tetrad $\hat{e(0)}_{\mu i}$
(minimal tetrad) playing the role to change coordinates between
both spaces and a true tensor $\hat{\gamma}_{\mu \nu}$ related to
displacement transformations, an idea which is implicit in the
expression (9) of the general metric tensor. 

\section{Minimal tetrads}

Assuming that the structure of the non linear tetrads is given by
equations (8) and (26), we are now interested in the behavior of
the theory when considered as an expansion in $\lambda$. To start
from the lower terms, we are going to call ``minimal tetrads" the
limit $\lambda\rightarrow0$ of the previously mentioned equations
(8) and (26) which are precisely the objects to be analyzed in this
section. However it is probably useful to briefly reassume
previously the standard gauge lagrangian formalism of gravity to
introduce some definitions and properties to be used in the
following. 

Once the non linear dynamical variables $\hat{e}_{\mu i}$ and
$\hat{A}_{\mu i j}$ have been obtained,  the ordinary Einstein
equations,  in terms of the metric tensor $g(\lambda)_{\mu\nu}$, can
be deduced from gravitational gauge lagrangian build up with the
Field--Strength Tensor $\hat{F}_{\mu\nu i
j}=\partial_{[\mu}\hat{A}_{\nu]i j}+\hat{A}_{[\mu i
k}\hat{A}_{\nu] j}^k$.  Taking advantage however of the knowledge
of its internal structure as given by equations (8), (23) or (26)
we are ready to reach a deeper insight into the meaning and
possibilities of the non linear approach. 

Nevertheless being the cosmological constant an essential feature
of the quantum vacuum we are going to include it as an ingredient
of the theory in the empty space.  The presence or not of a
cosmological term has, as we will see,  important consequences. 

To calculate the field equations we start from the gravitational
gauge lagrangian density written in the form:
\begin{equation}
L=\hat{e}\hat{e}\,(\lambda)^{\mu i}\hat{e}(\lambda)^{\nu
j}\hat{F}_{\mu\nu i j}+\hat{e}\Lambda
\end{equation}
where $\hat{e}$ is the determinant,  $\Lambda$ the cosmological
constant and $\hat{e}(\lambda)^{\mu i}$ the formal inverse
of $\hat{e}(\lambda)_{\mu i}$.  To explore the dependence on
$\lambda$ of the solutions we split the tetrad
$\hat{e}(\lambda)_{\mu i}$ as:
\begin{equation}
\hat{e}(\lambda)_{\mu i}=\hat{e}(0)_{\mu
i}+\lambda\hat{\gamma}_{\mu i}
\end{equation}
where according to (26)
\begin{equation}
\hat{e}(0)_{\mu
i}=\frac{1}{\hat{M}}(\partial_\mu\hat{\varphi}_i+\hat{A}_{\mu i
j}\hat{\varphi}^j)
\end{equation}
and $\hat{\gamma}{\mu i}$ is given by (28)

To take into account (32) and (28) suitable Lagrange multipliers
$\Delta^{\mu i}$ and $\Sigma^{\mu i}$ can be introduced.  So that
we add to (30) the following auxiliary term:
\begin{equation}
\Delta L=\hat{e}\Delta^{\mu i}\left[\hat{e}(o)_{\mu
i}-\frac{1}{\hat{M}}(\partial_\mu\hat{\varphi}_i+\hat{A}_{\mu i
j}\hat{\varphi}^j)\right]+\hat{e}\Sigma^{\mu i}\left[\hat\gamma_{\mu
i}-(\frac{1}{\hat{M}}\gamma^*_{\mu
i}+\frac{\hat\varphi_i\hat\varphi_j}{r^2}\gamma_{\mu^j})\right]
\end{equation}

Under all these assumptions the field equations can be written in
the following form:
\begin{equation}
\hat{F}(\lambda)_{\mu i}-\frac{1}{2}\hat{e}(\lambda)_{\mu
i}\hat{F}(\lambda)=\hat{e}(\lambda)_{\mu i}\Lambda
\end{equation}
\begin{equation}
\partial_\nu M^{\nu i\mu j}-M^{\mu [ i \nu k}A_{\nu k}^{j ]}=0
\end{equation}
with the addition of the conditions (28), (32) and being the Lagrange
multipliers equal to zero.  Here $\hat{F}_{\mu
i}=\hat{e}(\lambda)^{\nu j}\hat{F}_{\mu\nu i j}$ ,  $
\hat{F}=\hat{e}(\lambda)^{\mu i}\hat{F}_{\mu i}$ and $M^{\mu i \nu
j}=\hat{e}(\lambda)\hat{e}(\lambda)^{\mu [i}\hat{e}(\lambda)^{\nu
j]}$. 

The solution of (35) is highly simplified when $\hat{A}_{\mu i j}$
is redefined as follows:
\begin{equation}
 \hat{A}_{\mu i
j}=\hat{e}(\lambda)^\alpha_iD_\mu\hat{e}(\lambda)_{\alpha
j}+B_{\mu i j}, 
\end{equation}
where $D_\mu$ is the ordinary  Christophel covariant derivative
acting on the index $\alpha$ of the tetrad
$\hat{e}(\lambda)_{\alpha j}$.  Introducing the last in (35) we get
immediately $B_{\mu i j}=0$.  So that finally the motion equations
are given again by the conditions (32),  (28) and the more familiar
relations:
\begin{equation}
G(\lambda)_{\mu\nu}=\frac{1}{2}g(\lambda)_{\mu\nu}\Lambda
\end{equation}
\begin{equation}
\hat{A}_{\mu i
j}=\hat{e}(\lambda)_i^{\alpha}D_\mu\hat{e}(\lambda)_{\alpha j}
\Rightarrow{B_{\mu i j}}=0
\end{equation}
where $G(\lambda)_{\mu\nu}$ is the Einstein tensor written in
terms of the previously defined  metric $g(\lambda)_{\mu\nu}$
(9). Again for a vanishing $k$ the field equations becomes the
corresponding ones to the Poincar\'e group. 

We have employed Lagrange multipliers to reach a cleaner
result, however it is interesting to notice that the formal
variations with respect to $\varphi_i$ in the lagrangian (30)
formally leads to a motion equation which becomes the covariant
divergence of the Einstein tensor and therefore is satisfied
identically.  This result can be easily understood by considering
that,  as deduced from its transformation properties,  the fields
$\varphi_i$ are isomorphic to coordinates ( the cartesian ones in
the Poincar\'e case );  therefore they can be eliminated from the
theory by a simple choice of coordinates.  A result that allows us, 
coherently with the non linear approach,  to interpret them
dynamically as Goldstone bosons of the theory. We remark that, due
to the use of Lagrange multipliers, equation (32)is strictly, at
this level, a pure definition useful to discuss the structure of
the dynamical variables.

 As we have pointed out the minimal
structure having the formal properties of a tetrad is given by
$e(0)_{\mu i}$.  An important object as far as,  according to (9), 
it can be used to provide us with a basis suitable to express
gravity as an expansion in $\lambda$ .  Therefore the question is
now to calculate the formal expression and consequences of the
introduction of this minimal tetrad.   We are going to consider
however,  in the first place,  the Poincar\'e group,  briefly
reassuming the results obtained in a previous paper  \cite{Martin}, 
at this purpose we take the limit $\lambda\rightarrow0$ in
equation (8) which,  in this case, reads:
\begin{equation}
e(o)_{\mu i}=\partial_\mu\varphi_i+A_{\mu i j}\varphi^j
\end{equation}

Taking $\lambda=0$ in equation (36) and multiplying then by
$\varphi^j$ one gets:
\begin{equation}
A_{\mu i j}\varphi^j=e(0)^\alpha_iD(0)_\mu e(0)_{\alpha
j}\varphi^j + B_{\mu i}=e(0)^\alpha_iD(0)_\mu[e(0)_{\alpha
j}\varphi^j]-\partial_\mu\varphi_i + B_{\mu i}
\end{equation}
where $D(0)_\mu$ is the covariant Christophel derivative with
respect to the Greek index of the tetrad,  constructed with the
metric tensor $g(0)_{\mu\nu}$, and $B_{\mu i}=B_{\mu i j}
\varphi^j$.  When $k\rightarrow 0$ it follows from the definition
(40)
\begin{equation}
e(0)_{\alpha j}\varphi^j=\partial_\alpha\sigma
\end{equation}
with $\sigma=\frac{1}{2}\mu^2=\frac{1}{2}\varphi_i\varphi^i$. Using
(42) and (43) in (40) we obtain:
\begin{equation}
e(0)_{\mu i}=e(0)^{\alpha}_iD(0)_\mu D(0)_\alpha \sigma+B_{\mu i}. 
\end{equation}
Multiplying now by $e(0)^{i}_{\nu}$ we get finally the expression
of the minimal metric tensor which reads:
\begin{equation}
g(0)_{\mu \nu}=D(0)_\mu D(0)_\nu \sigma+B_{\mu \nu}, 
\end{equation}
that for the vacuum solutions , $B_{\mu\nu}=0$,  reduces to:
\begin{equation}
g(0)_{\mu\nu}=D(0)_\mu D(0)_\nu \sigma
\end{equation}

It can be easily seen that (46) is a particular case of the more
general one:
\begin{equation}
D_\mu D_\nu\sigma=\frac{1}{d} g_{\mu \nu} \Box \sigma
\end{equation}
being $d$ the dimensionality of the space. It must be emphasized
that here the question is not to find the structure of $\sigma$, that we know ``a priori", but to state that the existence of a
scalar density $\sigma$ satisfying (45) implies,  as an
integrability condition, the maximally symmetric character of the
space. Not surprisingly when equation (44) is concerned it can be
shown that the space becomes directly Minkowskian. This is again a
consequence of the role played by the fields $\varphi_i$ as
Goldstone bosons of the theory with respect to the translations. In
fact, choosing the $\varphi_i$ in (44) as coordinates we easily
verify that the metric adopt the pseudoeuclidean value $\eta_{i
j}$, valid only in the absence of the cosmological constant. To
summarize briefly the Poincar\'e case puts in evidence that the
non linear gauging of the group of isometries leads us to a
minimal tetrad which generates precisely the metric of a
pseudoeuclidean  background space, a result that support the idea
to link gravity to the translational connections. At the same time, from the limit $\lambda\rightarrow 0$ in equation (37), we realize that Poincar\'e group  excludes (at expected) the presence of a cosmological constant.

As a consequence space--time physics can be described by two
different dynamics, one related with the characteristic length
$\lambda$ is of course the ``true" gravitational interaction, while
that associated with the properties of the quantum vacuum, relevant in the limit $\lambda\rightarrow 0$, is
phenomenologically characterized by the cosmological constant
$\Lambda$. The necessity
 therefore to include $\Lambda$ in a realistic scheme strongly
suggests to study the group of isometries of a maximally symmetric
space. Thus we recover the explicit form of equation (29) which
becomes:
\begin{equation}
\hat{e}(0)_{\mu
i}=\frac{1}{\hat{M}}(\partial_\mu\hat{\varphi}_i+\hat{A}_{\mu i
j}\hat\varphi^j)
\end{equation}
following the same lines of the previous calculations we get:
\begin{equation}
\hat{e}(0)_{\mu
i}=\frac{1}{\hat{M}}\hat{e}(0)^{\alpha}_{i}D(0)_\mu\left[\frac{1}{\hat{M}}D(0)_\alpha
\hat{\sigma}\right]
\end{equation}
where $\hat{\sigma}=\eta^{i j}\hat{\varphi}_i\hat{\varphi}_j$

Now multiplying by $\hat{e}(0)_{\nu}^{i}$ we obtain finally
\begin{equation}
g(0)_{\mu \nu}=\frac{1}{\hat{M}^2}\left[D(0)_\mu
D(0)_\nu\hat{\sigma}-\frac{\hat{M'}}{\hat{M}}D(0)_\mu
\hat{\sigma}D(0)_\nu \hat{\sigma}\right]
\end{equation}
being $\hat{M'}$ the derivative of $\hat{M}$ with respect to
$\hat{\sigma}$. 

Equation (50) can be alternatively written in a more compact
expression of the form:
\begin{equation}
g(0)_{\mu \nu}=\frac{\mbox{Const}}{\hat{M}}D(0)_\mu D(0)_\nu \
\tilde{\sigma}
\end{equation}
where $\hat{\sigma}=F(\tilde{\sigma})\tilde{\sigma}$ and
$F(\tilde{\sigma})$ satisfy the differential equation:
\begin{equation}
F'(\tilde{\sigma})\tilde{\sigma}+F(\tilde{\sigma})=\mbox{Const}\,{\hat{M}}. 
\end{equation}
Taking now the constant equal to one, the trace of equation (51)
becomes:
\begin{equation}
\hat{M}=\frac{1}{4}\Box(0)\tilde{\sigma}
\end{equation}
which substituted in (51)leads us to:
\begin{equation}
D(0)_\mu D(0)_\nu\tilde{\sigma}=\frac{1}{4}g(0)_{\mu
\nu}\Box(0)\tilde{\sigma}, 
\end{equation}
in which we recognize equation (45) which implies, as an
integrability condition, that $g(0)_{\mu \nu}$ describes the metric
of a maximally symmetric space having, consequently,  a constant
curvature,  thus compatible with the presence of a cosmological
term in the theory. We remark that the presence of the factor
$\hat{M}^{-1}$ in equation (51) is essential to obtain a condition
like (45). The absence of this factor in the Poincar\'e case leads
us to (44) which implies the pseudoeuclidean character of the
space.

\section{Summary and conclusions}

As we have seen, equation (32) as well as its limit when $k$ tends
to zero are suitable structures to construct a gauge approach of
the Lorentz Group contained in a larger theory including
translations. To reach the integrability conditions of equations
(44) and (45) we have used a redefinition of the gauge field
$\hat{A}_{\mu i j}$ of the form:
\begin{equation}
\hat{A}_{\mu i
j}=\hat{e}(\lambda)^\alpha_iD_\mu\hat{e}(\lambda)_{\alpha
j}+B_{\mu i j}, 
\end{equation}
where the object $B_{\mu i j}$ has the meaning of a torsion
term, that in the presence of matter takes into account the
coupling of the  spin densities to the field $\hat{A}_{\mu i
j}$. In the absence of matter terms the field equations lead us to
the result $B_{\mu i j}=0$. So that as long as (44) and (49) have
been obtained in the absence of this contribution they are no
longer only a consequence of the definition of $\hat{e}(0)_{\mu
i}$ but a result of the vacuum motions equations. Notwithstanding
it must be emphasized the purely algebraic character of the motion
equation deduced from $\delta A_{\mu i j}$. It implies that,  when
only classical fields equations are concerned, $A_{\mu i j}$ can be
eliminated from the theory by substituting everywhere its value as
given by (36). As a consequence, once the expression for the minimal
metric tensor is fixed by integrability conditions, the only
relevant dynamical variable to describe pure gravity is given by
the translational connection $\gamma_{\mu i}$ or,  in other words, a
gauge theory of translations.

It is not in the scope of this paper to consider the introduction
of matter fields, limiting ourselves to the vacuum solutions in the
presence of a cosmological constant. It is immediate to see that
when we work with (53) the equation (49), for instance,  should be
substituted by the more general symmetric expression:
\begin{equation}
g(0)_{\mu\nu}=\frac{1}{\hat{M}}D(0)_\mu D(0)_\nu
\tilde{\sigma}+B_{\mu\nu}, 
\end{equation}
where $B_{\mu\nu}=\frac{1}{2}B_{(\mu i
j}\hat{\varphi}^j\hat{e}(0)_{\nu)}^i$.

It is worth mentioning that in the standard local field theory in
the presence of gravity $B_{\mu i j}$ stands for the matter sources
coupled linearly to $\hat{A}_{\mu i j}$,  (i.e.) the spin
densities. In a gauge field theory only fermions give rise to these
kind of contributions. It is known, on the other hand, that the
fermion spin densities $B_{\mu i j}$ are completely antisymmetric
when all indexes are reduced to have the same nature, therefore a
symmetric part of $B_{\mu\nu}$ should be absent in (54). In any
case this is a question to be considered in detail in the general
framework of a theory containing matter fields. 

The ``anomalous" dimensionality of the translational connection
$\Gamma(T)_i$ in equation (6) allows us to identify a natural
source for the introduction in the theory of a characteristic
length $\lambda$ intimately related to the gravitational forces. We
assume that $\lambda$ accomplish a fundamental role giving support
to a natural expansion for the gravitational
interaction. Nevertheless being $\lambda$ a dimensional
constant, some comments are probably in order. As we have pointed
out the idea of a characteristic length in gravitational physics
is commonly accepted by many authors and implicit in recent
developments. It can be easily understood by assuming that when the
scale of distances involved in a process are very large with
respect to $\lambda$, the relevant terms are the lower ones, the
results obtained when $\lambda\rightarrow 0$ widely support an
interpretation of this kind. Going however a little forward one
could consider $\lambda$ as a natural constant probably related
with the limit of validity of a geometrical description. 

The present scheme depends on two different constants that govern
the physics of the space-time. The first one, specially relevant
when $\lambda\rightarrow 0$, is the cosmological constant essential
to determine the choice of the gauge group concerned. The second
one is, of course,  $\lambda$ itself related to what we could call
properly gravitational forces. There appears, in this way, open
problems to be elucidated. From the point of view of the group
theory for instance, it is evidently interesting to try to
understand the meaning and behavior of displacement
transformations when a characteristic length like $\lambda$ is
present, an old problem on the other hand admiting different approaches. Concerning dynamics the close and foreseeable relationship
between $\lambda$ and the gravitational constant is another
obvious and relevant point. Both are questions to be considered on
the grounds of a general scheme including matter terms, work on
these aspects is actually in progress. 

\section*{Acknowledgements}
We acknowledge Prof. A.  Fern\'andez Ra\~nada and J.  Julve for useful discussions.  One of us (J.  Mart\'{\i}n) acknowledge financial support under the projects FIS2006-05319 of the Spanish MEC and SA010CO5 of the Junta de Castilla y  Le\'on.

\end{document}